\documentclass[a4paper]{sab}  

\usepackage{graphicx} 
\usepackage{txfonts,natbib} 
\bibpunct{(}{)}{;}{a}{}{,} 

\usepackage{pgf}

\usepackage[T1]{fontenc} 
\usepackage{t1enc} 
\usepackage{aeguill} 
\renewcommand\ackname{Acknowledgements.} 
\renewcommand\keywordname{Keywords.} 
\renewcommand\figureptname{Figure}
\renewcommand\tableptname{Table}

\renewcommand\noteaddname{Note added to proofs.}
\idline{Boletim da Sociedade Astron\^omica Brasileira, {\bf 29}, XX-XX}{1}
\begin{document} 
\title{Atmospheric spectroscopy at Gale Crater on Mars} 

\author{V. A. OLIVEIRA \inst{1} \and P. H. MOMBELLI} 
\institute{Univerisidade Federal do Pampa, Caçapava do Sul, Brazil \email{viniciusoliveira@unipampa.edu.br}}
\date{Received } 

\Abstract {Currently, Mars is the celestial object with the biggest quantity of devices made by mankind. This fact explain the astronomical quantity of data available about this planet, which allows several studies in several lines of research. Here we present the result of a monograph in Space Geophysics, in a graduation in Geophysics, the data used were collected by the device PFS (Planetary Fourier Spectrometer) in the mission MEX (Mars Express) launched in 2006. The main aim was the verification of the  existence of methane (CH$_4$) in a specific region in the atmosphere of the Red Planet. The target region, Gale Crater, is where the Rover Curiosity is right now. Therefore, the choice was because the strong evidence of liquid water in the past on this site, and because Curiosity has identified an unusual amount of methane in the local atmosphere.}{Atualmente, Marte é o objeto celeste com a maior quantidade de objetos feitos por humanos. Este fato explica a quantidade astronômica de dados disponíveis sobre este planeta, o que permite vários estudos em várias linhas de pesquisa.  No estudo atual, apresenta-se o resultado do Trabalho de Conclusão de Curso em Geociência Espacial, do curso de Geofísica, os dados usados foram coletados por um dispositivo PFS (Espectrômetro Planetário de Fourier) na missão MEX (Mars Express) lançada em 2006. O objetivo principal deste trabalho foi a verificação da existência de metano (CH$_4$) em uma região específica na atmosfera do Planeta Vermelho. A região-alvo, Cratera Gale, é onde o Rover Curiosity está. Sendo assim, a escolha foi devido a forte evidência de água líquida no passado deste local, e porque a Curiosity identificou uma quantidade incomum de metano na atmosfera do local.}


\maketitle 

\section{Introduction}
%
%

The term `GEO' has ceased to be exclusive to Earth studies, inclusive there is a new term: `Space Geoscience' to explain studies of celestial objects with detectable geological proprieties. In fact, there is increasingly non-stellar objects like moons, asteroids and exoplanets have been studied using Geology, besides to the first targets in the Solar System. 

This way, in 2015 was created the Laboratório de Geociências Espaciais e Astrofísica, with the purpose of encouraging and facilitating the studies in Spacial Geoscience. In an effort to start this initiative, we begin a spectroscopic study, in infrared, of the atmosphere of Mars. This study presents the direct objective to analyze and identify methane (CH$_4$) in the Martian atmosphere, first in a specific region and after over all the planet. In later stages the study intends to determine the abundance of this substance, since it is an organic tracer with fragile survival in atmospheres as that of Mars. This characteristic could be an indicative of a methane source active nowadays.

\section{Mars}

Mars is the fourth planet of the Solar System, from the Sun, its name is in honor of the god of war of Ancient Rome. Its reddish color is because the large quantity of powder of iron oxide III (Fe$_2$O$_3$), which cover almost all the planet \citep{christensen03}  Although in many aspects Mars is similar to Earth, the spatial dimension is smaller, especially the diameter (53.2\% of the Earth), the mass (10.7\% of the Earth) and the mean surface gravity, only 3.71 m/s$^2$ \citep{willians04}. 

Mars presents a rarefied atmosphere, being estimate to be 100 times less dense than Earth's \citep{nasa05}. The chemical composition is practically equal to terrestrial, but the abundances differ considerably. The main discrepancies, estimated by Owen (1992), are the CO$_2$ abundance of 95\% in the Mars' atmosphere, against 0.04\% in Earth; the N$_2$ about 2.7\% in Mars and 78\% in Earth; and the O$_2$ 0.13\% and 21\%, respectively. 

The time coordinates of Mars were adapted to the specific situation of the planet, then it was defined an unity named Solar Longitude (L$_S$). This unity is proportional to the angle between an imaginary line linked Mars and the Sun with the equinox line in Mars. Then, the Martian months correspond to the variation of L$_S$ of 30$^{\rm o}$, however each month has different quantities of Martian days (suns). These, in their turn, are equal to 24 hours and 34 minutes. At last, the Martian year is defined in 669 suns.

By contrast, the spacial coordinates system follows the same pattern as Earth, longitude and latitude, always referencing the planetary equator (zero latitude) and a central meridian (zero longitude). Whether the equator is easy to be identified, the central meridian must be choice arbitrary -- like did in Earth -- and, in 1972, the one chosen in that pass over a crater named Airy-0.

The main selection criterion to Gale Crater was the detection of unusual amounts of methane plumes by the Rover Curiosity in 2012. Considering that \citep{krasnopolsky04} estimated the lifetime of methane in the Martian atmosphere in approximately 340 terrestrial years. Therefore, the planet's atmosphere would not be able to maintain this element for a long period, this implies in a continuous production of CH$_4$. In fact, \citep{webster15} suggest that Mars is continually producing and releasing methane into the atmosphere.

\section{Observations}

We used the data collected with the Planetary Fourier Spectrometer (PFS), a device aboard the mission Mars Express (MEX) sent by European Space Agency (ESA) in partnership with {\it Agenzia Spaziale Italiana}. These data are available in the digital repository Planetary Science Archives (PSA).

\begin{figure}
 \label{fig:metano}
 \centering
  \pgfimage[height=5cm]{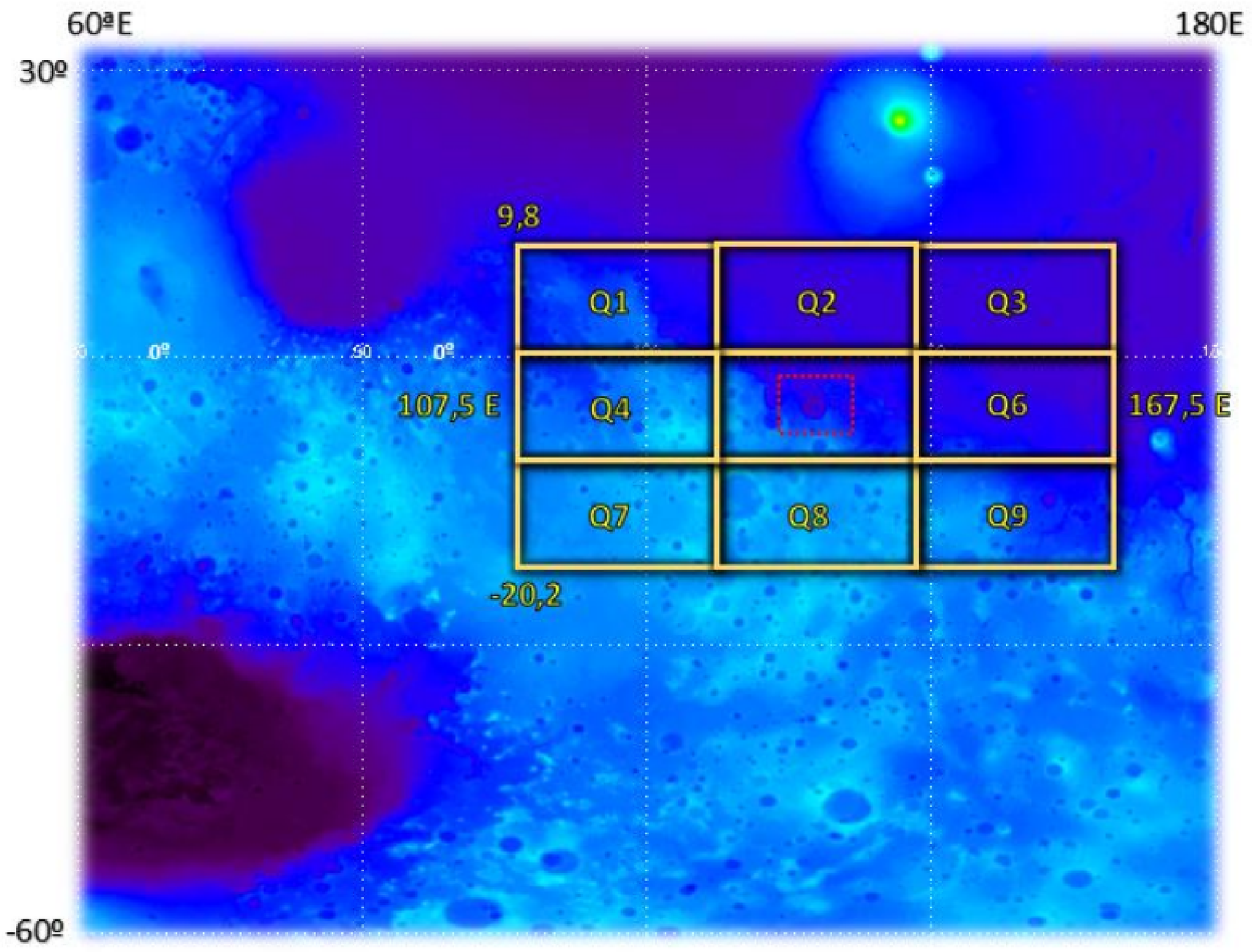}
  \caption{The division of the study area, Q5 indicates the Gale Crater, the figure image shows all the adjacent regions used in this work.} 
\end{figure}

According to \citep{formisiano04}, from all the observable CH$_4$ bands using PFS, only the 3\,000~cm$^{-1}$ band has a suitable SNR, in addition to having a prominent absorption peak of methane at 3\,018~cm$^{-1}$. For statistical and comparison purposes, it was chosen to work only with integrated spectrum over each region. The adjacent sections were chosen to verify the edge effects, since the Martian atmosphere is not static. The figure \ref{fig:metano} brings the visual information of all the chosen regions.

\section{Results}

Considering that the present study aims at a qualitative analysis, it was decided to estimate a regression curve in the section around the methane line in the spectrum for its correct identification. In this way it was possible to separate noise from true line, i.e., whenever the measured value exceeded a bias value the line was classified as true, otherwise it would just be noise. The cut-off is equivalent SNR measured (SNR$_{\rm med}$) greater than twice the SNR intrinsic (SNR$_{\rm int}$) of the equipment, in this case SNR$_{\rm int}$ is equal to 80 erg/s.cm$^3$.sr.cm$^{-1}$.

\begin{figure}
 \label{fig:results}
 \pgfimage[height=5.5cm]{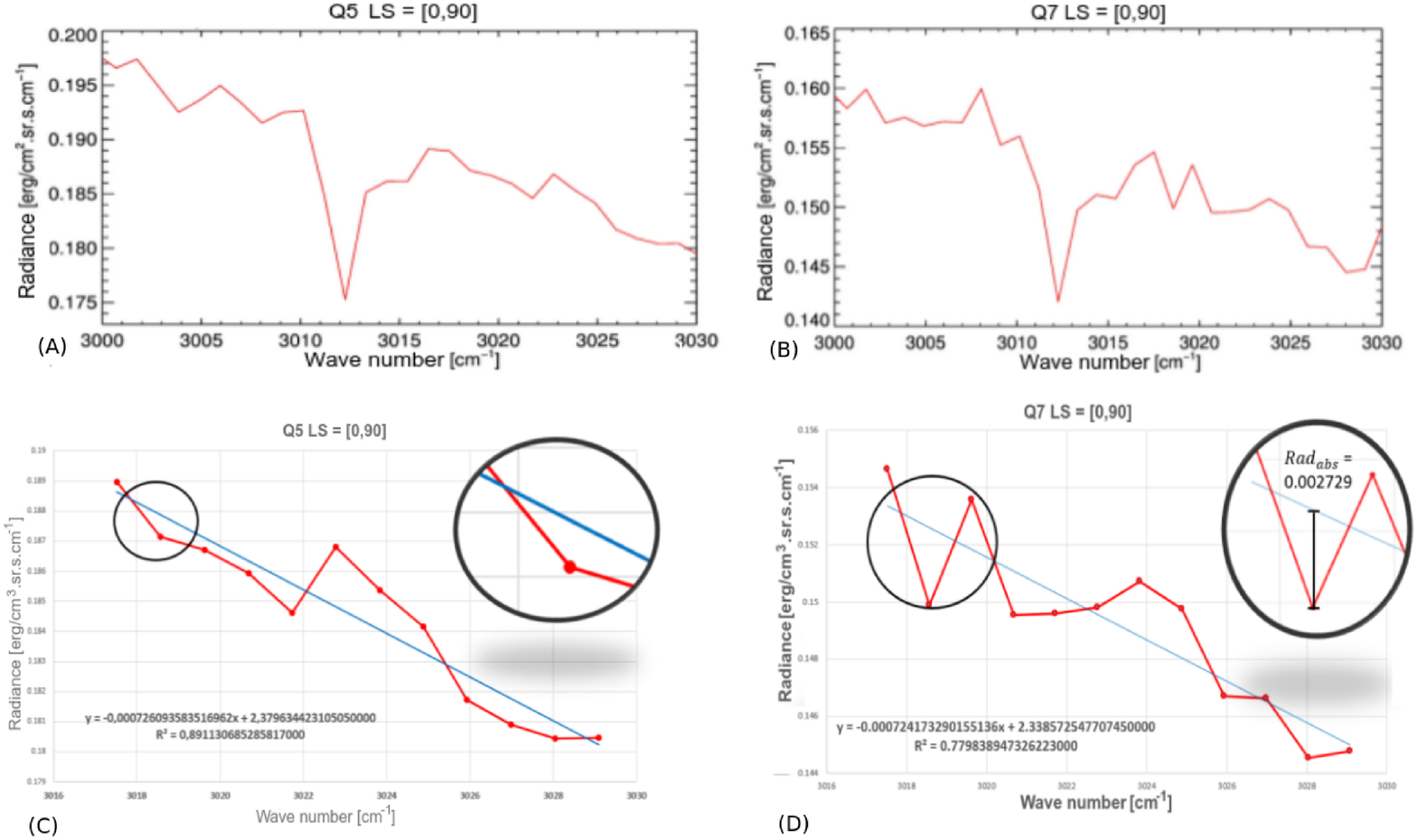}
 \caption{Integrate spectrum obtained (A) for the Q5 region and (B) for Q7 region. Qualitative result (C) in Q5 region and (D) in Q7 region, all of them during the Martian spring.} 
\end{figure}

Two of the results calculated by us are showed in the figure \ref{fig:results}, both are to Spring Season in Martian North Hemisphere; one over Gale Crater (Q5) and other to Q7 (our better measure). The other graphics are available, please demand the authors if you would to see.

The table 1 indicates the regions studied in which the presence of methane was identified, it is possible to note a seasonal variation of CH$_4$. 

\begin{table}
  \label{tab:resultados}
  \centering
  \caption{Qualitative results found in this work}
  \begin{tabular}{ccc}
   \hline
   Seasonal & Region with & SNR$_{\rm med}$ \\
   Period & indication of CH$_4$ & \\
   \hline
   Spring & Q1, Q3, Q7 & 2, 4, 3 \\
   Summer & -- & -- \\
   Fall & Q3, Q7 & 4, 2 \\
   Winter & Q1 & 2 \\
   \hline
  \end{tabular}
\end{table}

Germinale et al. (2011) indicated sections over Mars where it was detected methane, our results are in a good agreement with them, always during the spring. In fact, this season showed the best situation to determinate methane plumes on Mars. 

\section{Conclusion}

In any season studied in this work, there was no detectable amount of methane at Gale Crater (Q5), sometimes in sections around Q5, specially Q7 (the most promising of all them). A possible reason of this is that this study only used data from the Extended Mission 2 (EXT2) of MEX, from October 2007 to December 2009. It is worth pointing out that Curiosity does the cited detection of CH$_4$ only in 2014, five years after the data we have used was collected. Therefore, a more detailed analysis, over all the planet, could bring more information about the existence of methane in Mars. Likewise to use data from several mission, including many years, would be useful too.

\begin{acknowledgements} To MEX-M-PFS-2-EDR-NOMINAL-V1.0, a database available in the Planetary Data System (PDS); Dr. Giuseppe Sindoni and Dr. Marco Giuranna for the kindness of responding to the questioning emails, as well as giving some IDL scripts to start the analysis; and to the Department of Geophysics of the Universidade Federal do Pampa -- UNIPAMPA. \end{acknowledgements}


\begin{thebibliography}{} 
  \bibitem[CHISTENSEN et al. 2003]{christensen03} CHRISTENSEN, P.R. et al. 2003, Science Magazine, 300, 2056
  \bibitem[FORMISIANO et al. 2004]{formisiano04} FORMISIANO, V. et al. 2004, Science Magazine, 306, 1758
  \bibitem[GERMINALE et al. 2011]{germinale11} GERMINALE, A et al. 2011, Elsevier, 59, 137
  \bibitem[KRASNOPOLSKY et al. 2004]{krasnopolsky04} KRASNOPOLSKY, V. et al. 2004, Iacrus, 172, 537
  \bibitem[NASA 2005]{nasa05} NASA 2005, ``Exploration of the Planet Mars: Mars Facts''
  \bibitem[OWEN 1992]{owen92} OWEN, T 1992, Mars, 1, 818
  \bibitem[WEBSTER et al. 2015]{webster15} WEBSTER, C. et al. 2015, Science Magazine, 347, 415
  \bibitem[WILLIANS 2004]{willians04} WILLIANS, D.R 2004, NASA Goddard Space Flight Center, 1
\end{thebibliography}
\end{document}